\documentclass[lettersize,journal]{IEEEtran}
\usepackage{amsmath,amsfonts}
\usepackage{algorithmic}
\usepackage{algorithm}
\usepackage{array}
\usepackage[caption=false,font=normalsize,labelfont=sf,textfont=sf]{subfig}
\usepackage{textcomp}
\usepackage{stfloats}
\usepackage{url}
\usepackage{verbatim}
\usepackage{graphicx}
\usepackage{cite}
\usepackage[dvipsnames]{xcolor}
\begin{document}

\title{Differentiable Machine Learning-Based Modeling for Directly-Modulated Lasers}

\author{Sergio Hernandez Fernandez,~\IEEEmembership{Student Member,~IEEE}, Ognjen Jovanovic~\IEEEmembership{Member,~IEEE}, \\ Christophe Peucheret~\IEEEmembership{Member,~IEEE}, Francesco Da Ros~\IEEEmembership{Senior Member,~IEEE}, Darko Zibar
        % <-this % stops a space
\thanks{Sergio Hernandez, Ognjen Jovanovic, Francesco Da Ros and Darko Zibar are with the Department of Electrical and Photonic Engineering, Technical University of Denmark, 2800 Lyngby, Denmark (e-mail: shefe@dtu.dk, ognjo@dtu.dk; fdro@dtu.dk; dazi@dtu.dk).

Christophe Peucheret is with the University of Rennes, CNRS, FOTON - UMR6082, F-22305 Lannion, France (e-mail: christophe.peucheret@univ-rennes.fr)}% <-this % stops a space
\thanks{Manuscript received September 28, 2023; revised August 16, 2021.}}

\markboth{IEEE Photonics Technology Letters}%
{Hernandez \MakeLowercase{\textit{et al.}}: A Sample Article Using IEEEtran.cls for IEEE Journals}


\maketitle

\begin{abstract}
End-to-end learning has become a popular method for joint transmitter and receiver optimization in optical communication systems. Such approach may require a differentiable channel model, thus hindering the optimization of links based on directly modulated lasers (DMLs). This is due to the DML behavior in the large-signal regime, for which no analytical solution is available. In this paper, this problem is addressed by developing and comparing differentiable machine learning-based surrogate models. The models are quantitatively assessed in terms of root mean square error and training/testing time. Once the models are trained, the surrogates are then tested in a numerical equalization setup, resembling a practical end-to-end scenario. Based on the numerical investigation conducted, the convolutional attention transformer is shown to outperform the other models considered.

%All the models are assessed in terms of sample-to-sample accuracy and complexity, and their performance is compared directly to the response of the laser rate equations within offline optimization. Throughout the approaches studied, the Convolutional Attention Transformer shows resilient performance in a wide variety of waveforms and symbol rates, while maintaining high parallelization capabilities.
\end{abstract} % We need to change the processing time as the metric, but I currently dont know to what, so I will keep it in red.


\section{Introduction}
Directly-modulated lasers (DMLs) play a crucial role as part of intensity-modulation and direct-detection (IM/DD) systems in short-reach communication links. Due to their inherent simplicity, DMLs have the potential to achieve efficiency gains in both power consumption and cost-effectiveness compared to alternative transmitter technologies \cite{Huang2021BeyondApplications}. However, their modulation bandwidth (around 30 GHz at 25 °C) limits the symbol rate of commercial DMLs to the 50 Gbaud range, making them a less compelling option as Ethernet throughput requirements increase \cite{Yamaoka:23}. Apart from the ever-present phase and intensity noise, effects such as waveform distortion or frequency chirping dominate when pushing their modulation rate, hindering their potential in terms of transmission distance and data throughput. One can benefit from an increased modulation bandwidth by driving the laser with higher current values, at the cost of a lower extinction ratio and  degraded sensitivity. Finding an optimal balance between signal degradation mechanisms is therefore a complex task. This has lead to the investigation of different mitigation techniques to increase the symbol rate while maintaining a sufficient signal quality.

 Equalization (EQ) and pre-distortion have been broadly used in this context, as they force the received signal to resemble the original unaltered waveform. Nevertheless, previous equalization solutions have relied on the separate optimization of the transmitter and receiver, disregarding the potential gains obtained through their simultaneous optimization  \cite{Huang2021NonlinearSystems}. To achieve further data throughput improvements, joint optimization of the transmitter and receiver using end-to-end (E2E) learning has gained traction as an optimization approach for optical communication systems, pushing their performance closer to their theoretical capacity \cite{Karanov2020ConceptModel, Srinivasan:23}.

The conventional approach to gradient-based optimization in E2E learning is based on a differentiable channel model \cite{Karanov2020ConceptModel}. The DML small-signal response can be easily approximated by differentiable methods, at the cost of constraining the peak-to-peak amplitude of the modulated signal to impractically low levels in a realistic scenario. The more suitable large-signal dynamics are however governed by nonlinear differential rate equations, for which no analytical solution can be obtained \cite{Zhu2018DirectlyLasers}. This limitation poses challenges in achieving a differentiable channel. Although ordinary differential equation (ODE) solvers and optimization approaches (reinforcement learning, gradient-free) have been proposed as gradient estimators, they often require considerable computational overhead \cite{Yankov:22}. To enable E2E learning and facilitate the estimation of gradients within the communication system, a locally accurate DML model is required \cite{Wang2021TheSimulation}.

This letter builds upon our work in \cite{US:ECOC23} showcasing the application of data-driven optimization techniques to derive differentiable DML models. The model performance analysis is based on four different data-driven models, namely time-delay neural networks (TDNN), Volterra filters, long-short term memory (LSTM) and convolutional attention transformers (CATs).  In this work we integrate each model into a new system optimization setup and conduct a comparative analysis of the generated signals with the laser rate equation output. The objective is to evaluate the models' gradient estimation performance from a more contextualized perspective as part of a larger optimization system, instead of assessing them as mere function estimators. A quantitative comparison between the models is conducted in terms of normalized root mean square error (NRMSE) and train/test time, while providing a visual qualitative comparison through the use of eye diagrams. 

Comparing the different architectures, the results show that the CAT model can achieve improved NRMSE performance in training and testing throughout the analyzed symbol rates while maintaining  a GPU processing time comparable to its alternatives. CATs are therefore expected to offer an efficient solution for optimizing DML-based communication systems where the direct use of ODE solvers would be impractical. %The proposed Transformer method is compared to three other common function estimators in dynamical system analysis (Volterra series, time-delay neural networks (TDNNs) and LSTMs).

\section{Numerical Setup}
\subsection{Data-Driven Modeling}
{The main goal in data-driven modeling is to estimate the underlying dynamics of a system solely by means of its input-output relations. In the present case, this is achieved by using the laser rate equations \cite{Cartledge} to relate input current sequences to the envelope of the optical output waveforms.} {The use of the rate equations does not replace experimental validation, but it allows the assessment of models without the limitations of experimental setups (measurement noise, increased data acquisition time, lack of flexibility in the setup, etc.)}
%In the case of DML modeling we know that local context is key, i.e. recent samples provide more valuable data than older samples to the prediction task. This idea motivates the use of a model that captures such dependencies as part of its structure.
Alternatively to more established techniques in laser modeling, like circuit-level models \cite{Srinivasan:23}, 
we propose the utilization of CATs \cite{Li2019EnhancingForecasting}. CATs employ convolutions to model the dependencies within temporal sequences. This approach offers several advantages: (i) it restricts the utilization of past sequence samples in prediction, (ii) it captures waveform patterns rather than individual sample relations, and (iii) it has strong awareness of the order of the samples within the sequence. Although recurrent architectures are also based on temporal context, they need to calculate previous states sequentially in order to infer future samples. CATs break this bottleneck by processing the full time sequences at once, making better use of parallelization hardware and memory resources \cite{Li2019EnhancingForecasting}.

%Transformers are machine learning structures specifically designed for parallel processing of numerical sequences, eliminating the need for recurrent elements. This is achieved through the attention mechanism, which determines the relevance of each input element for the given task.

%Although the original Transformer \cite{Vaswani2017AttentionNeed} has shown poor performance in time series data due to its inability to capture temporal dependencies between sequential elements \cite{Zeng2022AreForecasting}, specialized structures have been proposed to overcome this limitation.

\begin{figure*}[t]
    \centering
    \includegraphics[width=0.925\linewidth]{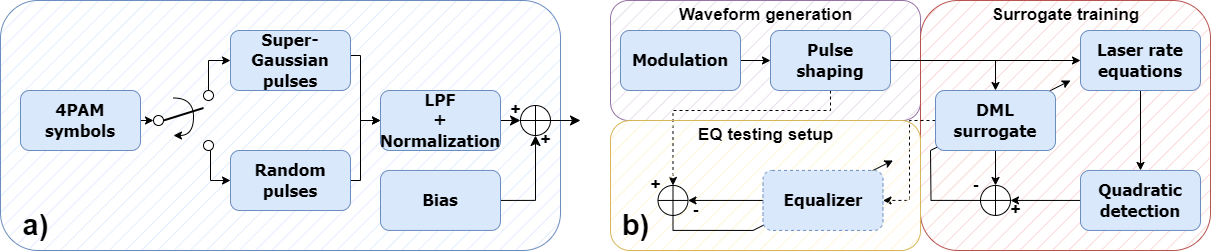}
    \caption{Block diagrams of a) data acquisition and b) model setup}
    \label{fig:eq_setup}
\end{figure*}

To maximize the accuracy of the data-driven model across various scenarios, the input data must encompass a wide range of waveforms and amplitudes, providing deep understanding of the laser's dynamic behaviour. %Ideally, the pulse shaping block generating the data should use few input parameters while yielding a large amount of different output waveforms to avoid overfitting. 
This is addressed, as shown in Fig.~\ref{fig:eq_setup}a, by alternating between two types of pulse shapes: super-Gaussian pulses and random pulses. The random pulses are sampled as vectors from a folded normal distribution $\mathcal{N}(0.5,1)$. The parameters for the super-Gaussian pulses, namely the temporal full width $e^{-0.5}$, $T_{0}$ and the order $n$ are sampled from a folded $\mathcal{N}(0.25T_\mathrm{sym},T_\mathrm{sym})$ and uniform $\mathcal{U}(1,6)$ distributions, respectively. $T_\mathrm{sym}$ is the symbol period, reciprocal of the symbol rate.
The amplitude of the pulses is modulated according to equiprobable pulse-amplitude modulation (4PAM) symbols. Subsequently, the pulses undergo min-max normalization and low-pass filtering (LPF) to prevent out-of-band leakage. The pulse shaping is randomized again every 8 symbols (with 32 samples per symbol) until completing a 1024-sample sequence of mixed pulse shapes. The training data set comprises $2^{13}$ sequences, totaling $2^{23}$ samples, while the validation set consists of $2^{17}$ samples. The symbol rate of the driving signal is varied to introduce different levels of  distortion, obtaining a distinct model for every symbol rate investigated. To solve the laser rate equations, a fifth-order Runge-Kutta (RK4,5) solver is utilized. As shown in Fig.~\ref{fig:eq_setup}b, the solution obtained from the solver serves as the ground truth for the surrogate models, establishing the relationship between the input modulation current (generated waveform) and the laser output after quadratic detection (optical power).

\begin{table}[t]
  \centering
\caption{Model hyperparameters used} \label{tab:params}
\begin{tabular}{|c|c|c|c|}
        \hline  & \textbf{CAT}  & \textbf{TDNN} & \textbf{LSTM}\\
        \hline  Number of hidden nodes & 128 & 2048 & 64\\
        \hline Number of hidden layers & 2 & 1 & 2 \\ 
        \hline Activation function & ReLU & ReLU & ReLU \\
        \hline  Number of MLP sublayers & 2 & 2 & -\\
        \hline  Convolutional window length & 9 & 31 & -\\
        \hline  Embedding vector size & 128 & - & -\\
        \hline Number of attention heads & 8 & - &- \\ 
        \hline {Training peak GPU mem. usage (GB)} & {9.28} & {0.89} & {1.82} \\
        \hline
\end{tabular}
\end{table}%

The proposed CAT model adopts a decoder-only structure, consisting of three main blocks: learned positional embeddings (LPEs), {full} convolutional attention sublayers {(omitting sparsity due to decreased training performance)}, and 2-layer multilayer perceptrons (MLP) with ReLU hidden activation. 
%The network incorporates residual connections based on the RK2 ordinary differential equation (ODE) Transformer structure \cite{Li2021ODETranslation}, and each sublayer output undergoes layer normalization.
A linear layer is employed to reduce the dimensionality of the hidden features. For comparison purposes, three additional models were investigated: a second-order Volterra filter with 16-sample memory, a TDNN, and an LSTM. The specific values for each network hyperparameter are summarized in Table~\ref{tab:params}.

\subsection{Equalization setup}
To demonstrate the proof-of-concept, all the trained surrogate models were evaluated within a numerical back-to-back transmission setup {(optical phase not considered)} with a receiver equalizer. %Thus, their potential to achieve link gains in a real optimization environment can be showcased.
A simple FIR-based equalizer trained on NRMSE was selected as testing scenario, as its simplicity allows to focus on the accuracy of the DML model. It is important to note that the optimality of the equalizer structure for this task is not a primary concern within the present scope, as the focus lies on the predictive potential of the surrogate models. The equalization task is performed on a per-sample basis, using square-pulse-shaped 4PAM symbols. It should be emphasized that none of the surrogate models was trained specifically on pure square waveforms, ensuring a fair assessment of their inference capabilities. As depicted in Fig.~\ref{fig:eq_setup}b, the loss is calculated by calculating the NRMSE between the low-pass-filtered waveforms at the input of the DML and the signal at the output of the equalizer. Since the studied surrogates cannot perfectly replicate the laser rate equations, the learned equalizer coefficients are also tested on the output waveforms from the ODE solver to gain insight into the extent of metric distortion induced by the models. The comparison between the losses obtained from the surrogate models and the rate equation serves as the primary benchmark, as it reflects the generalization capabilities of each model in handling a previously unseen scenario.

\section{Numerical results}
\subsection{Surrogate models}
The numerical simulations performed can be divided into surrogate and equalizer optimization. The DML response exhibits spectral characteristics that cause waveform distortion to increase as the symbol rate $R_s$ increases, especially beyond the relaxation frequency $f_R$. The high-rate region close to $f_R$ is therefore of particular interest. Within the surrogate training, we evaluated models across six distinct symbol rates, spanning the range $0.1 f_R$ to $1.2 f_R$. In each instance, we used an Adam optimizer with recommended decay values ($\beta_1 = 0.9, \beta_2 = 0.999$), along with min-max normalized mean square error (NMSE) as loss metric, later converted into its squared root form, NRMSE, for unit matching purposes. As a proof of concept, laser phase and intensity noise were not considered for simplicity. Each surrogate model was trained for 400 epochs, selecting the one achieving the lowest test loss. The optimal hyperparameters for each model were obtained through grid search. The selected models were then utilized for the equalization task, so their potential as part of a link optimization setup could be verified.

 \begin{figure}[t]
   \centering
    \includegraphics[trim={0.25cm 0 1.6cm 1.52cm},clip, width=0.9\linewidth]{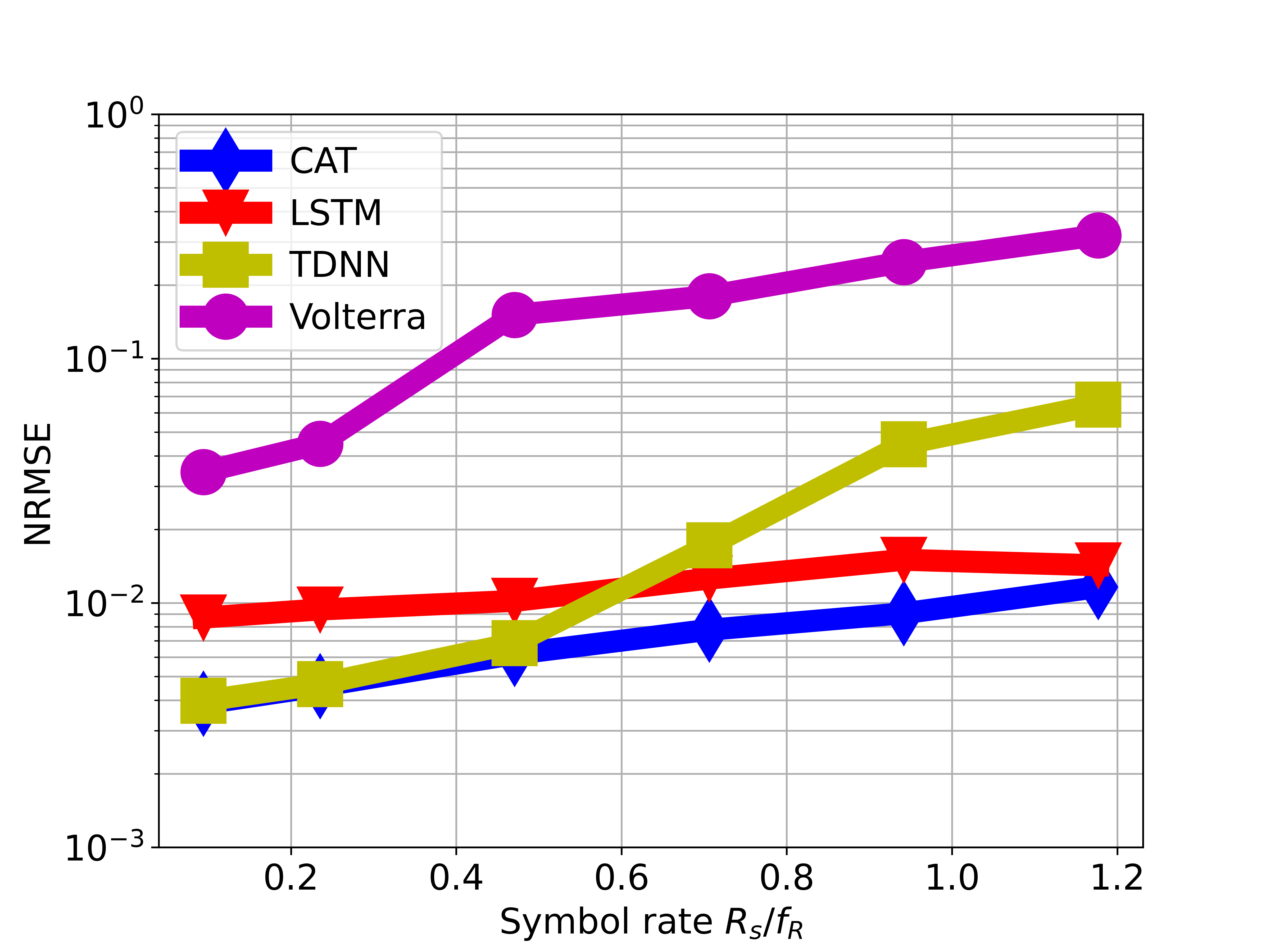}
    \caption{NRMSE scores of the studied models}
    \label{fig:rmse}
    \vspace{4.8pt}
\end{figure}

\begin{figure}[t]
    \centering
    \includegraphics[trim={0.25cm 0.1cm 1.6cm 1.52cm},clip, width=0.9\linewidth]{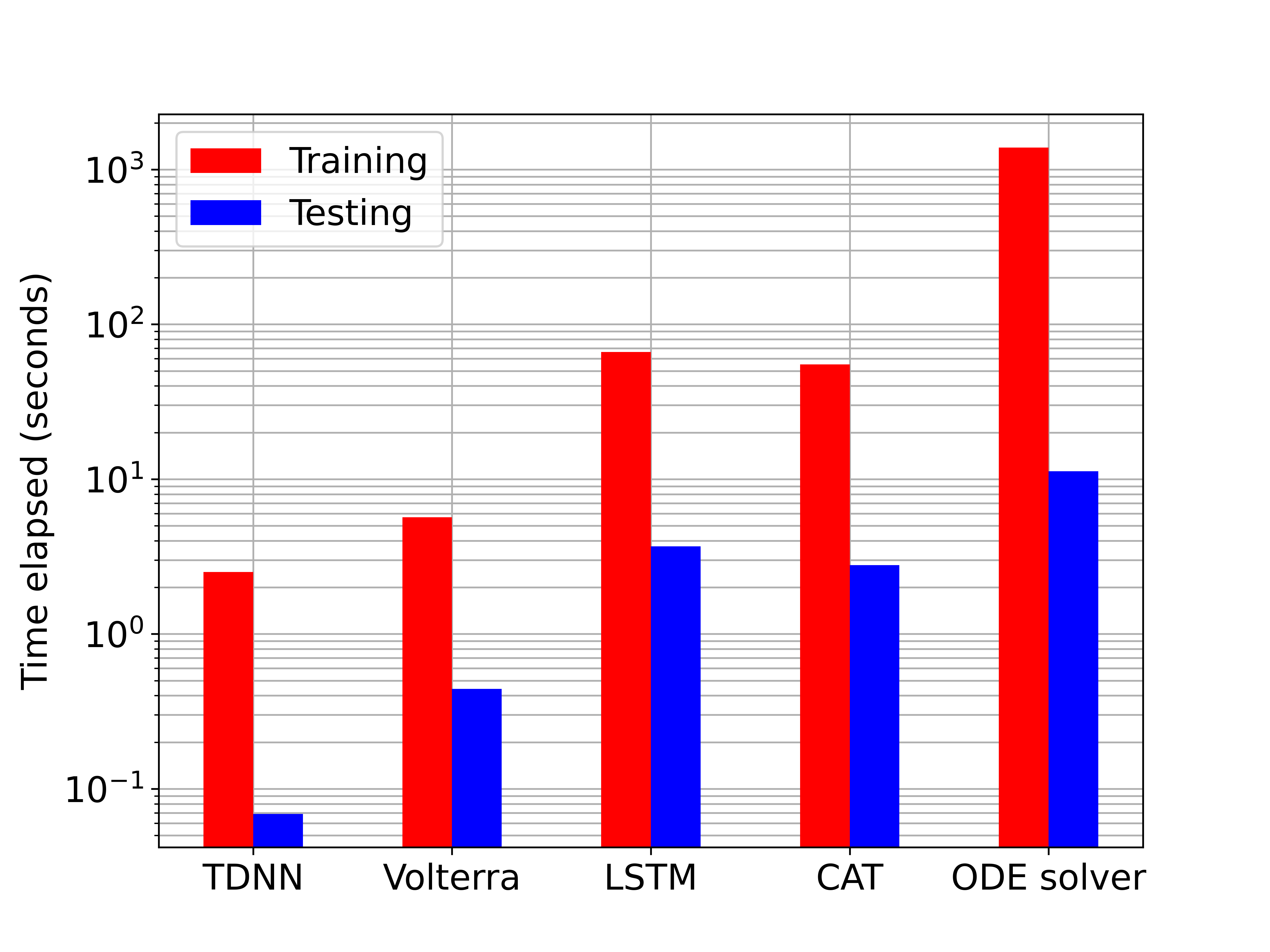}
    \caption{Time elapsed (per epoch) by the presented models}
    \label{fig:times}
    \medskip
\end{figure}

\begin{figure*}[t]
   \centering
    \includegraphics[width=1\linewidth]{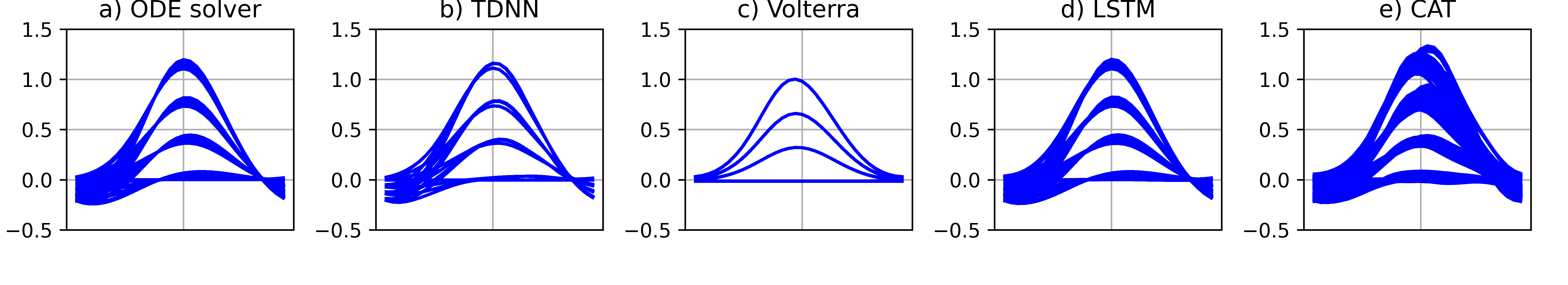}
    \caption{Eye diagram of a 4PAM Gaussian pulse train at $R_s \approx f_R$ for a) ODE solver, b) TDNN, c) Volterra filter, d) LSTM and e) CAT. }
    \label{fig:eye_diag}
\end{figure*}

The most straightforward validation in a supervised time series inference model is to compare its predictions with the desired sequences sample by sample. This approach is shown in Fig.~\ref{fig:rmse}, where the NRMSE value is plotted as a function of the symbol rate. Over the symbol rates analyzed, most of the models deliver similar performance, although the CAT seems to deliver better performance than its peers and falls below the $ 10^{-2}$ mark. This is especially true for higher symbol rates, where only the LSTM is able to approach its performance. The Volterra-based model is the main outlier, with an error over 10\% in the high rate region. Despite the substantial differences between the models, all the loss curves hint at the correlation between $R_s$ and the waveform distortion introduced, showing that as the symbol duration becomes shorter, the output sequences are more difficult to match. The TDNN showcases this tendency, showing relatively good NRMSE figures at low $R_s$ that worsen gradually as the symbol rate is increased. Further context is given in Fig.~\ref{fig:times}, where the average processing time for a train and test epoch on the utilized NVIDIA A100 GPU is compared. The ODE solver score shows the elapsed time for the generation of the target sequences. The figure pictures the importance of the model architecture in the inference speed of the networks: although the CAT has significantly more training parameters than the other surrogates, it operates at comparable times, even outperforming the LSTM. It is also clear that all of the proposed models add substantial time savings compared to the ODE solver. Another useful insight can be obtained by looking at Fig.~\ref{fig:eye_diag}, where the response of each model to a train of 4PAM Gaussian pulses is represented in the shape of eye diagrams. The output of the ODE solver to the same signal was added as a reference. Although all 4 models show reasonable convergence compared to the ODE case, there are some noticeable outliers. While the Volterra filter and the TDNN seem to oversimplify the DML dynamics compared to the ODE solver, the CAT is more sensitive to small changes in position and amplitude in samples. The former effect is probably due to the relatively low number of training parameters in the models, while the latter may be due to the one-to-many mapping in the positional encoding of the CAT. However, this drawback may be less relevant in real scenarios where noisy input data will affect the results of the output waveform to some extent. Even if the eye diagrams give a good intuition of the behaviour of each model, they show the response to a very specific input, while the NRMSE scores yield a broader analysis throughout the different waveforms.
%\textit{Although all 4 models show reasonable convergence compared to ODE cases (with the exception of TDNN), Volterra filters and LSTMs have consistent performance through the symbol $2 ^{ 10}$ shown, while CAT seems more sensitive to small changes in position and amplitude in samples. This may be due to positional encoding in the model that changes input to the network based on the position of the sample, although its value remains constant. However, this drawback may be less relevant in real scenarios where noisy input data will affect the results of the output waveform to some extent. In terms of capturing the real width of the output pulse, the CAT is slightly better able to track than its counterparts and tends to shorten the modulated pulse duration.}

\subsection{Equalizer optimization}

\begin{figure}[t]
    \centering
    \includegraphics[width=1.0\linewidth]{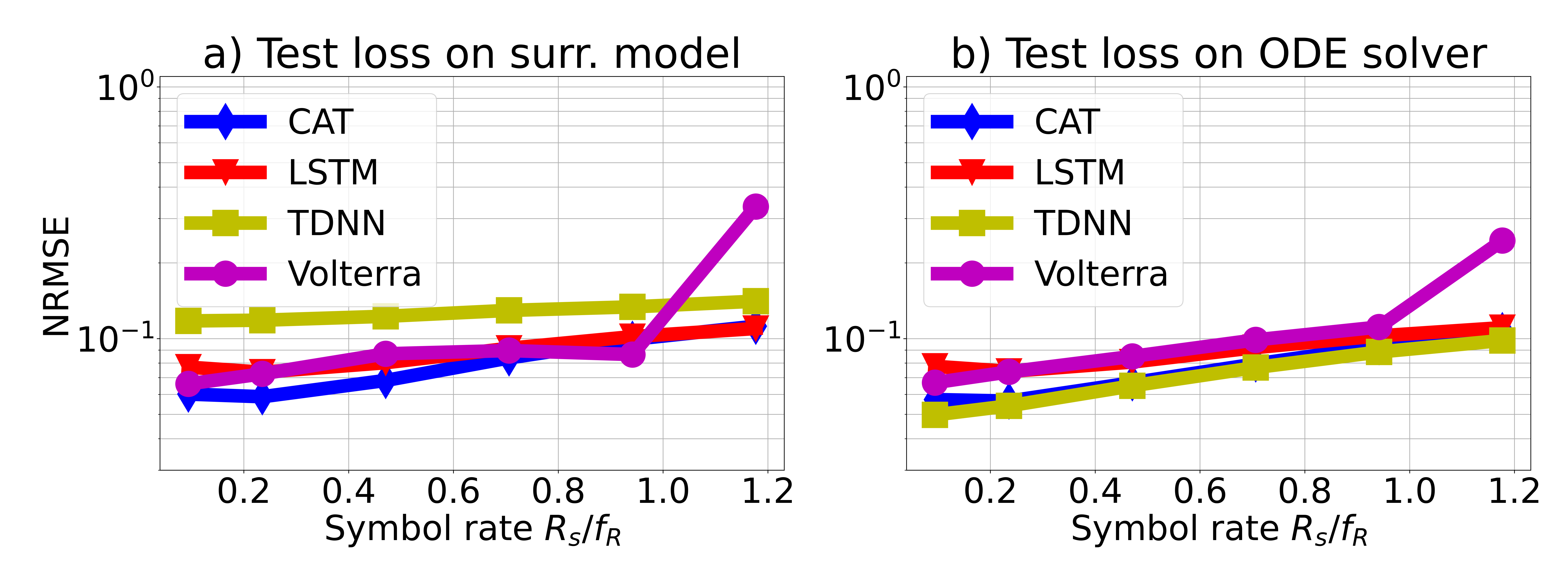}
    \caption{Equalization NRMSE tested on a) surrogate models b) ODE solver}
    \label{fig:eq_mse}
\end{figure}

Additional insights can be extracted from the equalization setup, where the MSE between the received and transmitted waveforms was obtained. The FIR equalizer is based on a trainable 31-tap filter, using random 4PAM transmitted symbols as a data source. In all cases, the symbol sequences utilized are identical for fair comparison. {It must be noted that regardless of the utilized model, all the simulations are based on noiseless numerical data, yielding a simplified approximation of the behavior of a physical DML. The equalization performance presented may therefore not translate fully to an experimental setup, especially when compensating for linear distortion.} Fig.~\ref{fig:eq_mse} establishes a comparison between the loss calculated based on the response of each of the models and the loss obtained when using the ODE solver as estimator of the laser response. %It becomes apparent that, even though all of the surrogates yield similar overall performance, the difference between the two plots is significant in certain cases.
While the LSTM and the CAT show almost identical curves in each case, the TDNN and Volterra MSE loss is significantly poorer when tested on themselves than on the ODE solver. This could be due to waveform artifacts (hinted in the eye diagrams) that distort the signal only when the testing is performed on the model, but make the equalizer more robust towards impairments during training. {This effect could however hinder the performance of the model as part of a larger E2E setup, as the compensation could focus on dynamics that do not correspond to the behavior of a physical laser.} The Volterra filter seems to deliver a relatively solid response in the low-rate region, but it progressively degrades when approaching higher symbol rates. Even if the TDNN delivers the best ODE-based testing performance, its scores differ noticeably from the self-testing case, making it potentially unreliable as estimator of the DML response.

\section{Conclusions}
This study has proposed a series of differentiable surrogate models for directly modulated laser links. In addition to the usual loss metrics, the models were tested in an equalizer-based optimization setup to showcase their prospects in a real setting. %The analysis shows the complexity of choosing a model that resembles the laser response under every scenario and the variety of factors that must be taken into account.
Throughout the metrics obtained, the convolutional attention transformer has shown high resilience to different waveforms and symbol rates, while maintaining relatively low inference times thanks to its parallelization capabilities. Our results show the potential of data-driven models as faster substitutes for ODE solvers and derivative-free gradient approximators in the context of link optimization. 
%-------------------------------------------------- Acknowledgements Section -------------------------------------------------------%

\section{Acknowledgements}
This work was financially supported by the ERC-CoG FRECOM project (no. 771878) and the Villum YIP OPTIC-AI project (no. 29334).

\bibliographystyle{IEEEtran}
\bibliography{references.bib}

\newpage

\vfill

\end{document}